\documentclass[12pt]{article}
\begin{document}
\baselineskip = 1.5\baselineskip
\title{Remarks on Bessel beams, signals and superluminality}
\author{J. T. Lunardi\footnote{On leave from Universidade Estadual de Ponta Grossa,
Setor de Ci\^encias Exatas e Naturais, Departamento de
Matem\'atica e Estat\'{\i}stica.
Ponta Grossa, PR, Brazil.} \footnote{e-mail: \textit{lunardi@ift.unesp.br}} \vspace{.1cm}\\
Instituto de F\'{\i}sica Te\'{o}rica\\
Universidade Estadual Paulista\\
R. Pamplona, 145\\
01405-900. S\~{a}o Paulo - SP\\
Brazil}

\maketitle

\begin{abstract}
We address the question about the velocity of signals carried by
Bessel beams wave packets propagating in vacuum and having well
defined wavefronts in time. We find that this problem is
analogous to that of propagation of usual plane wave packets
within dispersive media and conclude that the signal velocity can
not be superluminal.
\end{abstract}

PACS numbers: 02.30.Nw; 03.50.De; 04.30.Nk

Keywords: Wave propagation; Bessel beams; signal velocity;
superluminality.

\newpage

\section{Introduction}

It is generally accepted that the causality principle in special
relativity imposes the speed $c$ of light in vacuum as the upper
limit for the velocity of propagation of signals or interactions.
By the way, Maxwell equations in material media allows, in
special circumstances such as anomalous dispersion, plane wave
solutions propagating with superluminal (faster than $c$)
\textit{phase} velocities \cite{Jackson}. Although these
velocities cannot be readily identified with the signal velocity,
the question about the possibility of building a superluminal
signal by superimposing these solutions naturally arises.

The above problem was studied by Sommerfeld and Brillouin at the
beginning of the last century \cite{SB}. In order to define
precisely a velocity of propagation Sommerfeld considered only
signals carried by wave packets which have \textit{well defined
wavefronts} (and possibly ends) in time, which we call \textit{SB
signals}. The signal itself, or its ``main part", independently
of the way it is defined, is obviously confined to the region
behind the wavefront. These packets were assumed to be normally
incident on a semi-infinite dispersive medium, in which boundary
the arrivals of both the wavefront and the signal were assumed to
coincide. The Sommerfeld's main result was that, irrespective the
material medium, the wavefront of such a packet propagates always
with the speed $c$. Consequently, the arrival of the signal at a
given point inside the medium can occur only after that (or
simultaneously, if the medium is nondispersive) this point is
reached by the wavefront. Therefore, Sommerfeld result implies
that the speed $c$ of light in vacuum is an upper bound to the
velocity of propagation of SB signals. Brillouin studied in great
detail the evolution of these wave packets within dispersive
media, giving a precise definition for the signal velocity, which
agrees with the Sommerfeld result. In what concerns the group
velocity, which can be greater than $c$ in the presence of
anomalous dispersion, or even negative, Brillouin stated that
``...the group velocity has a meaning only so long as it agrees
with the signal velocity."

Recently, there is a growing interest on the subject of
superluminal wave motion in the literature. For example, we can
quote experiences measuring superluminal velocities in the
passage of wave packets through barriers (see, for example,
\cite{Wang}-\cite{Nimtz} and references therein) and a
considerable amount of works, both in theoretical and
experimental views, concerning the superluminality of Bessel
beams and X waves (\cite{Saari97}-\cite{Sauter2} and references
therein). These works raised questions about the interpretation
of such superluminal velocities, specially in what concerns the
meaning of \textit{signal} velocity and its connection with the
\textit{group} velocity and the causality principle. So, this is
still a very debated subject.

In this letter we are concerned to Bessel beams wave packets, from
which X waves are a special case. These beams are (inhomogeneous)
plane wave solutions of the homogeneous scalar wave equation in
vacuum which propagate with superluminal \textit{phase}
velocities. X waves are localized waves built up as special
superpositions of Bessel beams and propagate rigidly (without
dispersion) in vacuum with superluminal velocities. From these
nondispersive properties some authors suggested that this
superluminality could also be associated with the signal velocity
\cite{Recami, MRR, Mugnai}. We analyze the question of the maximum
velocity of signals carried by Bessel beams wave packets. To this
aim  we follow the approach of Sommerfeld and Brillouin cited
above and consider only SB signals, i.e., those carried by Bessel
beams wave packets having well defined wavefronts (and ends) in
time. We first study the \textit{chopped Bessel beam}, which have
a finite duration at its source. From the analysis of this packet
we identify a mathematical analogy between its propagation
properties and those of usual (inhomogeneous) plane wave packets
propagating within a dispersive medium, namely a tenuous
electronic plasma. This is our main result, because such analogy
makes possible a straightforward application of Sommerfeld result
to conclude that \textit{the wavefronts of these packets
propagate with the speed $c$, while the wave packets distort
while propagating}. This is a curious result because we are
dealing with propagation of waves \textit{in vacuum}. As a direct
consequence, the velocity of the signals carried by these wave
packets can never be superluminal, independently of the way it is
defined.

We also consider briefly the experiment of Mugnai, Ranfagni and
Ruggeri \cite{MRR}, which posed a question about the
superluminality of signals carried by Bessel beams. An explanation
for the measured velocities in this experiment was given in
references \cite{RodriguesCausal,Sauter2} in terms of
interference phenomena, showing that the superluminal velocities
of the peaks moving along $z$ axis were not causally connected
and, therefore, did not represent signal velocities. To be able
to say something about the signal velocity in this experiment we
argue that the waves produced in it can be viewed as a kind of
\textit{finite aperture approximation to chopped X waves}. The
last ones are ideal waves (they need an infinite aperture to be
produced) built up as linear superpositions of chopped Bessel
beams. We show that chopped X waves have wavefronts moving with
velocity $c$ and thus the SB signals carried by them can not have
superluminal velocities. Also, we suggest that the observed
superluminal peaks can be qualitatively explained from the fact
that the chopped X wave distorts along the propagation, showing a
kind of \textit{reshaping phenomenon}, as observed in
\cite{RodriguesCausal}.

The generalization of the analogy with dispersive media to a large
class of superpositions of Bessel beams is straightforward and
will be done in Section 4. In the last section we make our
concluding remarks. In particular, we comment on some
discrepancies between our results and others in the literature
concerning the wavefront velocities of chopped Bessel beams and
chopped X waves.

\section{Bessel beams wave packets}

Bessel beams are cylindrically symmetric solutions of the scalar
homogeneous wave equation in vacuum. They are given by
\cite{Durnin, Saari97, Recami, RodriguesCausal}
\begin{equation}
\Psi^{(k_{\rho},\omega)}(\rho,z,t)=J_0(\rho k_{\rho})\exp\{i(k_z
z-\omega t)\}\, ,\label{Bessel}
\end{equation}
where $J_0(x)$ is the Bessel function of the first kind and order
zero and the parameters $k_z$, $k_{\rho}$ and $\omega$ satisfy the
following relation
\begin{equation}
k_z^{2}=\frac{1}{c^{2}}\left[\omega^{2}-c^2 k_\rho^{2}\right]\,
.\label{dispersao}
\end{equation}
Thus, from a mathematical point of view, \textit{any two}
parameters among $k_z$, $k_\rho$, and $\omega$ can be chosen
independently. For the purposes of this letter we choose $k_\rho$
and $\omega$ as the independent ones and assume they are real.
Making so, $k_z$ will be given by (\ref{dispersao}) and it can,
in principle, be imaginary.

At this point we emphasize that the above relation is
mathematically identical to the dispersion relation of a tenuous
electronic plasma if we identify $c^2 k_\rho^2$ with the square
of the plasma frequency $\omega_p^2$ \cite{Jackson}.

If $\omega^{2}\geq c^2 k_\rho^{2}$ the Bessel beam (\ref{Bessel})
represents an unidimensional wave motion propagating along the
$z$ direction, whose propagation properties are governed by the
real wave number $k_z$. The surfaces of constant phase are planes
perpendicular to the $z$ axis which propagate with the phase
velocity $v_p$, given by
\begin{equation}
v_p =\frac{\omega}{k_z}\, ,\label{vfase}
\end{equation}
such that $|v_p|\geq c$, i.e., the phase velocity in vacuum is
superluminal. These waves are also inhomogeneous plane waves, due
to the presence of Bessel function $J_0(\rho k_\rho)$ in
(\ref{Bessel}). In the case $k_\rho=0$ the Bessel beams
degenerate to usual homogeneous plane waves, with $|v_p|=c$. On
the other hand, if $\omega^{2}< c^2 k_\rho^{2}$ the wave number
$k_z$ is imaginary. In this case there is no wave propagation in
the $z$ direction, but instead of a behavior analogous to
absorption or attenuation.

The most general (complex) solution formed from superposition of
Bessel beams (\ref{Bessel}) is given by
\begin{equation}
\Psi(\rho,z,t)=\Psi^{+}(\rho,z,t)+\Psi^{-}(\rho,z,t)\label{supkrw}\,
,
\end{equation}
where
\begin{equation}
\Psi^{\pm}(\rho,z,t)=\int_{0}^{\infty}dk_\rho
\int_{-\infty}^{\infty}d\omega \, A^{\pm}(k_\rho,\omega)  J_0(\rho
k_{\rho})\exp\{i[\pm k_z(k_\rho,\omega) z-\omega t]\}
\label{lr}\, .
\end{equation}
In this expression $A^{\pm}(k_\rho,\omega)$ are spectral
distributions and the dispersion relation is given by
\cite{Sauter}
\begin{eqnarray}
k_z(k_\rho,\omega)=\left\{
\begin{array}{ll}
\frac{\omega}{c}\sqrt{1-\frac{c^2 k_\rho^{2}}{\omega^2}},&{\mathrm{if }}
\,\,|\omega|\geq ck_\rho\\
\frac{i|\omega|}{c}\sqrt{\frac{c^2 k_\rho^{2}}{\omega^2}-1},&{\mathrm{if }}
\,\,|\omega|< ck_\rho\\
\end{array}
\right. \, .\label{dispersao2}
\end{eqnarray}
With this choice of signals the terms $\Psi^{+}(\rho,z,t)$ and $
\Psi^{-}(\rho,z,t)$ in the superposition (\ref{supkrw})
correspond, respectively, to right and left moving wave packets.

Specifying the boundary conditions at $z=0$,
\begin{eqnarray}
\Psi(\rho,0,t)&=&\Psi^{0}(\rho,t)
\label{contg1}\, ;\\
\frac{\partial\Psi}{\partial z}(\rho,0,t)
              &=&\Psi^{'0}(\rho,t)\, ,
\label{contg2}
\end{eqnarray}
we can determine the spectral distributions
$A^{\pm}(k_\rho,\omega)$ in (\ref{lr})\footnote{To derive this
result we make use of orthogonality of Bessel functions
$$
\int_{0}^{\infty}d\rho \rho J_0(\rho k_\rho)J_0(\rho
k'_\rho)=\frac{1}{k_\rho'}\delta(k_\rho -k'_\rho)\, .
$$}
\begin{equation}
A^{\pm}(k_\rho,\omega)=\frac{k_\rho}{2\pi}\int_{-\infty}^{\infty}dt
{\mathrm{e}}^{i\omega t}\int_{0}^{\infty}d\rho \, \rho J_0(\rho
k_\rho)\left\{\Psi^{0}(\rho,t)\pm \frac{1}{ik_z(k_\rho,\omega)}
\Psi^{'0}(\rho,t) \right\} \label{A}\\
\, .
\end{equation}

To end this section we consider the situation in which only
propagating modes of Bessel beams (\ref{Bessel}) enter into
superposition (\ref{supkrw}), i.e., we restrict $ck_\rho<
|\omega|$ in (\ref{lr}). In this case we can write
\begin{equation}
k_\rho=\frac{|\omega|}{c}\sin \theta \quad {\mathrm{and}}\quad
k_z=\frac{\omega}{c}\sin \theta\, , \quad{\mathrm{with}}\quad
\theta\in \left[0,\frac{\pi}{2}\right)\, .\nonumber
\end{equation}
Using these relations we can change the integration variables in
(\ref{lr}) from $\left\{k_\rho,\omega\right\}$ to
$\left\{\theta,\overline{\omega}\right\}$ according to
\begin{eqnarray}
\overline{\omega}&=&\omega\, ; \nonumber\\
\theta&=&\arcsin\left(\frac{ck_\rho}{|\omega|}\right)\, .\nonumber
\end{eqnarray}
In this way the two terms in superposition (\ref{supkrw}) can now
be written as\footnote{We are taking into account the fact that
$J_0$ is an even function.}
\begin{equation}
\Psi^{\pm}(\rho,z,t)=\int_{0}^{\frac{\pi}{2}}d\theta
\int_{-\infty}^{\infty}d\overline{\omega}\,
B^{\pm}(\theta,\overline{\omega})
J_0\left(\rho\frac{\overline{\omega}}{c}
\sin\theta\right)\,{\mathrm{e}}^{i\frac{\overline{\omega}}{c}\left(\pm
z\cos\theta -ct\right)} \label{lrtw}\, ,
\end{equation}
where $B^{\pm}(\theta,\overline{\omega})$ are the spectral
distributions with respect to the new parameters. The parameter
$\theta$ is called the \textit{axicon angle}. Clearly the
superposition (\ref{supkrw}-\ref{lr}) is more general and
includes this last one.

When the axicon angle $\theta$ is fixed to some value $\theta_0$,
the last expression yields
\begin{equation}
\Psi^{\pm}_{\theta_0}(\rho,z,t)=\int_{-\infty}^{\infty}d\overline{\omega}\,
S^\pm(\overline{\omega}) J_0\left(\rho\frac{\overline{\omega}}{c}
\sin\theta_0\right){\mathrm{e}}^{i\frac{\overline{\omega}}{c}\left[\pm
z\cos\theta_0 -ct\right]}\, .\label{Xwaveint}
\end{equation}
This expression is frequently found in the literature and defines
the so-called X waves.\footnote{If
$S^+(\overline{\omega})=\delta(\overline{\omega}-\omega_0)$ then
$\Psi^{+}_{\theta_0}$ defines a Bessel beam (\ref{Bessel})
characterized by the axicon angle $\theta_0$ and frequency
$\omega_0$ propagating to the right. The same occurs if
$S^-=\delta$, with motion to the left.} Superposition
(\ref{Xwaveint}) can be viewed as a cylindrically symmetric
superposition of plane wave packets tilted over the $z$ axis by
an angle $\theta_0$ \cite{Saari97, Saari2000}. The particular
form of these packets is given by the spectral distributions
$S^{\pm}(\overline{\omega})$. As an example, if we choose
\begin{equation}
S^\pm(\overline{\omega})=\frac{-i{\mathrm{e}}^{i\left(\overline{\omega}-\omega_c
\right)t_0}}{4\pi\left(\overline{\omega}-\omega_c\right)}
\left[{\mathrm{e}}^{i\left(\overline{\omega}-\omega_c
\right)T_0}-1\right]\left[1\pm
\frac{\omega_c}{\overline{\omega}}\right]\, ,\label{Somega}
\end{equation}
then the plane wave packets are given by a rectangular pulse
modulation of a carrier of frequency $\omega_c$. In this
expression $T$ is the time duration of the pulses and $t_0$ is
the instant of time in which their wavefronts reach the origin.

At this point we observe that X waves (\ref{Xwaveint}) are
infinitely extended along the direction of propagation $z$. So,
they do not have a well defined wavefront into this direction and
thus do not define SB signals. In the next section we shall
consider the superposition (\ref{supkrw}-\ref{lr}) which ``chops"
(perpendicularly to $z$ direction) a Bessel beam characterized by
a given frequency $\omega_0$ and axicon angle $\theta_0$. From
these chopped Bessel beams we will construct the corresponding
chopped X waves.

\section{Chopped Bessel beams}

We now consider a SB signal consisting of a Bessel beam modulated
in time by a rectangular pulse at the plane $z=0$, i.e., a chopped
Bessel beam. This signal is introduced in \cite{RodriguesCausal}
through the boundary conditions (\ref{contg1}-\ref{contg2}), with
\begin{eqnarray}
\Psi^{0}(\rho,t)&=&{\mathrm{T}}(t)
J_0\left(\rho\frac{\omega_0}{c}\sin\theta_0\right)
                 {\mathrm{e}}^{-i\omega_0 t} \label{cont1}\\
\Psi^{'0}(\rho,t)&=&{\mathrm{T}}(t)\left(i\frac{\omega_0}{c}
                 \cos\theta_0\right)J_0\left(\rho\frac{\omega_0}{c}\sin\theta_0\right)
                 {\mathrm{e}}^{-i\omega_0 t}\, ,\label{cont2}
\end{eqnarray}
where $\omega_0$ and $\theta_0$ are \textit{fixed} constants and
the time modulation is given by
$\mathrm{T}(t)={\mathrm{\Theta}}(t)-{\mathrm{\Theta}}(t-\tau)$,
with ${\Theta}(t)$ being the Heaviside step function.
To determine the spectral distributions in the general
superposition (\ref{supkrw}-\ref{lr}) which correspond to these
boundary conditions we substitute these last expressions into
(\ref{A}). Then we have
\begin{equation}
A^{\pm}(k_\rho,\omega)=\delta\left(k_\rho-\frac{\omega_0}{c}\sin\theta_0\right)
C^{\pm}(\omega)\label{solvA}\, ,
\end{equation}
where
\begin{equation}
C^{\pm}(\omega)=\frac{1}{4\pi}\frac{{\mathrm{e}}^{i(\omega-\omega_0)\tau}-1}{i(\omega-\omega_0)}
\left\{1\pm\frac{\frac{\omega_0}{c}\cos\theta_0}{k_z(\frac{\omega_0}{c}\sin\theta_0,\omega)}
\right\}\, .
\end{equation}
We see that the spectral distributions $A^{\pm}(k_\rho,\omega)$
split into a factor depending only on $k_\rho$ and another factor
depending only on $\omega$. Then, we can write the right moving
term in (\ref{supkrw}) as
\begin{equation}
\Psi^{+}(\rho,z,t)=\int_{0}^{\infty}dk_\rho
\delta\left(k_\rho-\frac{\omega_0}{c}\sin\theta_0\right)\phi^{+}(k_\rho,\rho,z,t)
\label{dir} \, ,
\end{equation}
where
\begin{equation}
\phi^{+}(k_\rho,\rho,z,t)=\int_{-\infty}^{\infty} d\omega\,
C^{+}(\omega)J_0(\rho k_\rho) \exp\{i[k_z(k_\rho,\omega) z-\omega
t]\}\, .\label{disp}
\end{equation}
We can readily identify this last integral with that describing a
wave packet formed from superposition of (inhomogeneous) usual
plane waves and moving within a tenuous electronic plasma, with
plasma frequency given by $\omega_p=ck_\rho$. As the spectral
distribution $C^{+}(\omega)$ is analytical in the upper half of
the complex $\omega$-plane, this packet itself has a well defined
wavefront in time at $z=0$ \cite{Jackson,SB}. Therefore, we can
readily apply the results of Sommerfeld and Brillouin theory to
conclude that \textit{this wavefront moves with the velocity $c$
and the wave packet distorts while propagating}.

To completely describe the wave motion to the right, we must now
consider the $k_\rho$ integral in (\ref{dir}). So, we have
\begin{equation}
\Psi^{+}(\rho,z,t)=\phi^{+}\left(\frac{\omega^{0}_p}{c},\rho,z,t
\right)\, ,
\end{equation}
where in this expression $\omega^{0}_p=\omega_0 \sin\theta_0$.
Thus, all the conclusions after equation (\ref{disp}) remains
valid substituting the plasma frequency $ck_\rho$ by
$\omega^0_{p}$. These conclusions also hold for the left moving
part of the wave packet (\ref{supkrw}).

Superposing now the chopped Bessel beams above \textit{via}
spectral distributions $S^{\pm}(\omega_0)$, we have
\begin{equation}
\Psi^{\mathrm{chop}}_{\theta_0}(\rho,z,t)=
\Psi^{(+)\mathrm{chop}}_{\theta_0}(\rho,z,t)+
\Psi^{(-)\mathrm{chop}}_{\theta_0}(\rho,z,t)\, , \label{chopX}
\end{equation}
where
\begin{equation}
\Psi^{(\pm)\mathrm{chop}}_{\theta_0}(\rho,z,t)=\int_{-\infty}^{\infty}d\omega_0
\, S^\pm(\omega_0)\phi^{\pm}\left(\frac{\omega^{0}_p}{c},\rho,z,t
\right) \label{chopXlr}.
\end{equation}
It is easy to verify that in the absence of the time modulation
$T(t)$ this superposition would satisfy the same boundary
conditions as the X wave (\ref{Xwaveint}) if we identify the
spectral distributions in the two expressions. Accordingly, we
call superposition (\ref{chopX}-\ref{chopXlr}) a \textit{chopped
X wave}. As this wave is a linear superposition of chopped Bessel
beams, which have well defined wavefronts propagating with the
velocity $c$, also the chopped X waves will have this property.
As a consequence, they cannot carry superluminal SB signals.

Now we consider briefly the experiment of Mugnai, Ranfagni, and
Ruggeri, in which the authors measured superluminal velocities in
the propagation of X waves produced experimentally by a finite
aperture device \cite{MRR}. They posed the question about the
possibility of interpretation of these superluminal velocities as
velocities of signals. Causal explanations for the results of
this experiment were given in references \cite{RodriguesCausal,
Sauter2} based on simple models showing interference phenomena of
waves produced outside the axis along which the superluminal
velocities were measured. These explanations refute the
possibility of interpretating the measured superluminal
velocities as velocities of signals in that experiment. In the
reference \cite{RodriguesCausal} the authors call the attention
to the fact that this experiment shows a kind of generalized
\textit{reshaping phenomenon} occurring in free space,
characterized by the fact that the peak travels along the symmetry
axis faster than the wavefront.

In order to apply our analysis to this experiment we assume that
\textit{each pulse of the experimentally produced waves can be
viewed as a finite aperture approximation to an ideal chopped X
wave}. We consider the propagation in the region $z>0$, where the
plane $z=0$ contains the borders of the mirror which produce the
wave during the time interval $[0,\tau]$. To set the theoretical
(infinite aperture) model to the experiment we consider only the
right moving packet in (\ref{chopXlr}), with the spectral
distribution $S^+$ given by (\ref{Somega}). Now we interpret the
parameter $\omega_c$ as the frequency of the microwave carrier in
the experiment, $T$ as the time duration of the rectangular
modulation of this carrier and $t_0=\frac{R\sin\theta_0}{c}$ as
the instant in which the peak (which travels along the $z$ axis)
begins to be generated at the $z=0$ plane. $R$ is the radius of
the aperture and $\theta_0$ is the axicon angle fixed by the
experiment.\footnote{The axicon angle is given by
$\theta_0=\arctan \frac{d}{2f}$, where $d$ is the mean diameter
of the slit and $f$ is the focal length of the mirror.} In order
to produce the appropriate boundary conditions in the region of
finite aperture ($\rho\leq R$) at the plane $z=0$, the time
duration $\tau$ of the chopped X wave must satisfy $\tau\geq T
+2t_0$.

Assuming this model, our results about propagation of a chopped X
wave imply that its wavefront propagate with velocity $c$ and can
not carry a superluminal signal, a result that agrees with
references \cite{RodriguesCausal, Sauter2}. Also, we suggest that
the cited \textit{reshaping} phenomenon could be at least
qualitatively explained from the fact that the wave packet
distorts along its propagation.

\section{General results}

In the last section the essential feature allowing us to identify
the analogy between the propagation of the chopped Bessel beam in
vacuum and the propagation of usual plane wave packets within a
dispersive medium was the factorization of the spectral
distribution $A^{\pm}(k_\rho,\omega)$ into a product of a
distribution depending only on $k_\rho$ and another depending
only on $\omega$. We can generalize our results for the class of
all spectral distributions satisfying this property. For
simplicity, we will concern us to the right moving part of the
wave packet (\ref{supkrw}). All the conclusions will also hold
for the left moving part.

Let $A^{+}(k_\rho,\omega)=A_\rho(k_\rho)A_\omega(\omega)$. Then,
from (\ref{lr}) we have
\begin{equation}
\Psi^{+}(\rho,z,t)=\int_{0}^{\infty}dk_\rho\, A_\rho(k_\rho)
\phi^{+}(k_\rho,\rho,z,t)\, , \label{supkrwsep}
\end{equation}
where
\begin{equation}
\phi^{+}(k_\rho,\rho,z,t)=J_0(\rho k_{\rho})
\int_{-\infty}^{\infty} d\omega \, A_\omega(\omega)
\exp\{i[k_z(k_\rho,\omega) z-\omega t]\}\, .
\end{equation}
Following the same lines of the last section, we identify this
expression as representing an inhomogeneous usual plane wave
packet propagating within a tenuous electronic plasma, with
plasma frequency $ck_\rho$. Again, the $k_\rho$ integral in
(\ref{supkrwsep})  denotes a superposition of these packets. If
each component packet have a well defined wavefront in time at
some plane perpendicular to the $z$ axis, the complete
superposition will also have a time wavefront which, by our
previous analysis, propagates with velocity $c$.

To summarize, the velocity $c$ is the upper bound for the velocity
of any SB signal carried by a wave packet belonging to this class.

\section{Concluding remarks}

In this letter we studied the wavefront velocity of time limited
(chopped) Bessel beams wave packets and showed that they
propagate in vacuum in a way analogous to usual (inhomogeneous)
plane waves propagating within a tenuous electronic plasma. From
this analogy we were able to apply the Sommerfeld and Brillouin
results, originally conceiving the propagation of usual plane wave
packets within dispersive media, to conclude that the wavefronts
of these wave packets move always with velocity $c$, while the
waveform distorts along the propagation. These conclusions were
generalized for a large class of wave packets having well defined
wavefronts and described by factorizable spectral distributions.

The above results contradict some conclusions of references
\cite{RodriguesCausal, RodriguesCom}, in which the authors
conclude that chopped Bessel beams propagate \textit{without
distortion} with superluminal velocity. By a careful computation
of the spectral distributions in (\ref{lr})\textit{ which
correctly give the boundary conditions }(\ref{cont1}-\ref{cont2})
we found that the spectral distribution for the parameter
$k_\rho$ is a delta distribution which fixes this parameter at
the value $\frac{\omega_0}{c}\sin\theta_0$, contrary to the claims
of the authors of \cite{RodriguesCausal}, which consider $k_\rho$
as a parameter depending on the varying frequency $\omega$ (not
to be confused with the fixed parameter $\omega_0$,
characterizing the boundary conditions)\footnote{$k_\rho$, in our
notation, corresponds to the separation constant $\Omega$, in
their notation.}.

From superposition of chopped Bessel beams we constructed chopped
X waves and suggested that these waves could be viewed as
\textit{infinite aperture theoretical models} for the waves
produced in the experiment of Mugnai \textit{et al}, which posed
a question about the superluminality of signals carried by Bessel
beams. By our analysis the chopped X waves have wavefronts which
move with velocity $c$. Therefore, they can not carry
superluminal signals. Also, from the fact that these packets
distort along the propagation, we suggest that the
\textit{reshaping} phenomenon cited in \cite{RodriguesCausal}
could be at least qualitatively explained. It would be
interesting to develop the analogy with dispersive media further
in order to compare also quantitatively the results arising from
this model with the experimental data of Mugnai \textit{et al} and
with the models based on spherical waves presented in
\cite{RodriguesCausal, Sauter2}.

Our analysis introduced the formal analogy with dispersive media
as an alternative way to approach the problem of velocities of
signals carried by Bessel beams. We hope it can help us to better
understand this debated question.

\section{Acknowledgements}

I thank Prof. Dr. B.M. Pimentel and Dr. L.A. Manzoni for critical
reading the manuscript and CAPES/PICDT for partial support. I also
thank the anonymous referee for useful criticisms and suggestions.

\end{document}